\documentclass[12pt]{article}
\usepackage{graphics}
\usepackage{epsfig}
\usepackage{epstopdf}
\usepackage{amsmath}
\usepackage{array}
\usepackage[spaces,hyphens]{url}
\usepackage[colorlinks,allcolors=blue]{hyperref}
\usepackage{xcolor}
\begin{document}
	\begin{center}
		{\LARGE Emergence of cosmic
			space and the horizon thermodynamics in the braneworld scenarios\\[0.2in]}
		{ P. B.  Krishna, Adithya P. S. and Titus K. Mathew\\
			e-mail:krishnapb@cusat.ac.in, adithyaps@pg.cusat.ac.in, titus@cusat.ac.in \\ Department of
			Physics, Cochin University of Science and Technology, Kochi, India.}
	\end{center}

\abstract{ Expansion of the universe is caused by the departure from the holographic equipartition. 
	This principle, the law of emergence, first postulated in the context of Einstein's gravity  
	has been extended successfully to more general gravity theories like Gauss-Bonnet and Lovelock gravity.
	We derive the law of emergence for braneworld models of gravity, starting from the more fundamental and well established principle, the first law of thermodynamics. More specifically, we derive the law of emergence in the context of RS II braneworld, Warped DGP model and Gauss-Bonnet braneworld and compare the derived law with the one proposed by Sheykhi for the braneworld models. 
	We further show that the law of emergence leads to the maximization of horizon entropy in all these braneworld models. While the law of emergence effectively implies the maximization of horizon entropy, it could be derived from the first law of thermodynamics. Our results suggest that the horizon thermodynamics is the backbone of the law of emergence in the braneworld scenarios.

	\section{Introduction}
		In recent decades, much attention has been raised in exploring the intriguing connection between gravity and thermodynamics. This began with the pioneering work of Bekenstein and Hawking \cite{Hawking,medved2004conceptual,bekenstein1973black} in the context of black holes. These works disclosed that, 
	black holes possess
	a temperature directly proportional to the surface gravity at its horizon, 
	entropy
	proportional to the area of their horizon and can emit radiations like a black body. 
	Following these, the four laws of black hole mechanics have been proposed, which were indeed analogous to thermodynamic laws obeyed by ordinary macroscopic systems\cite{bardeen}. Incidentally it was suspected that, this thermodynamic characteristics is a unique property of black hole spacetime.
	But later, it was found that, thermodynamical behavior is a general feature of any spacetime, provided it is spherically symmetric \cite{eling2006nonequilibrium}. 
	In strengthening this result further, Jacobson derived Einstein’s field equations from Clausius relation, $\delta Q = TdS,$ by projecting them 
	to the horizon of Rindler observer,
	where $T$ is the temperature of the horizon, also 
	called Unruh temperature, $\delta Q$ is the change in the energy flux across the horizon\cite{jacob} and $ dS$ is the corresponding change in the entropy. Since then 
	many studies have been conducted, which further sheds light on the profound connection between gravity and thermodynamics \cite{thermodynical_aspects,pad_cosmos,Sindoni,Emergent_gravity}. Building on this foundation, Verlinde derived Newton's gravitational law by treating gravity as an entropic force \cite{Verli}. Later Padmanabhan derived gravitational law using equipartition law of energy and the thermodynamic relation $S = \frac{E}{2T}$, where $E$ is the effective gravitational mass, $S$ is the thermodynamical entropy and $T$ is the temperature of the horizon\cite{pad_Equi}.

	Relation of gravity with thermodynamics leads to deeper understanding. Thermodynamics is an
	emergent macroscopic phenomenon, since its features such as temperature, pressure etc, are inherently macroscopic and have no direct meaning 
	in the microscopic domain, instead they emerged out of the dynamics of the underlying degrees of freedom, the atoms or molecules. 
	Similarly, Einstein's theory of gravity could be an emergent 
	macroscopic phenomenon,  
	where its variables like curvature and the metric 
	have 
	no significance 
	at the microscopic level, instead they could be arised as the thermodynamic limit of some underlying degrees of freedom of the spacetime. 
	Padmanabhan explored this emergent nature of gravity\cite{padmanabhan2016exploringnaturegravity} and more importantly took a further bold step to propose that the cosmic space itself could 
	be 
	emerged 
	from some pre-existing internal degrees of freedom of the space, 
	which he termed as the ``atoms of space" \cite{padmanabhan2016atoms}. 
	It is very difficult to consider time as being emerged from some fundamental entity. However, in 
	cosmology, cosmic time is considered to be a unique universal time, which is the same 
	for all the fundamental co-moving observers. Hence time can be bifurcated from space and hence the idea, that the emergence of cosmic space is viable.  
	
	In 
	the emergent space approach, the expansion of the universe is considered to be driven by the difference or say discrepancy, $N_{surf} - N_{bulk},$  
	where $N_{bulk}$ is the number of degrees of freedom of the gravitating matter residing within the Hubble sphere and $N_{surf}$ is the spacial degrees of freedom on its boundary. The discrepancy, $N_{surf} - N_{bulk},$ between the degrees of freedom will decrease as the universe expands. This discrepancy will become zero, such that, 
	$N_{surf} = N_{bulk},$ as the universe reached its equilibrium state, the de Sitter epoch. This condition of the equality 
	is 
	termed as the holographic equipartition condition. Hence, in short, the expansion of the universe is caused by the departure from the holographic equipartition. The strength of this approach is that, this will naturally generate the Friedmann equations of the expanding universe, which are otherwise derived by substituting the Friedmann metric into the Einstein's equations of gravity. The method developed by Padmanabhan in this regard for Einstein's gravity has been extended by 
	Cai 
	to (n+1) dimensional Gauss-Bonnet and Lovelock theories for a flat universe\cite{cai2017gravitational}. It is further extended to non-flat universe by Sheykhi \cite{PhysRevD.88.067303}. This principle is also generalized to a variety gravity theories. 
	For more investigations on the law of emergence, see\cite{PhysRevD.96.063513,PhysRevD.99.023535,krishna2022emergence,krishna2020does, krishna2024emergence,vt2022emergence,hassan2023unified, sheykhi2009generalized,sheykhi2007thermodynamical,komatsu2017cosmological,komatsu2023evolution,komatsu2024cosmological,heydarzade2016emergent, volovick}. 
	
	It is very important to note that, these generalizations of the principle of the emergence of cosmic space or say, law of emergence are mere ad-hoc proposals, which are then ratified by deriving the respective Friedmann equations. It will be worth if this law could de derived from a more fundamental and well established principle. There are successful attempts to derive the expansion law from the  modified first law of thermodynamics. In \cite{Mirza}, the expansion law is derived from the first law of thermodynamics for a Kodama observer. In references \cite{mahith2018expansion,hareesh2019first}, authors derived the expansion law from the unified first law of thermodynamics in the context of Einstein, Gauss-Bonnet and Lovelock gravity theories.


	In the present work, we 
	derive the law of emergence from the first law of thermodynamics in the braneworld models of gravity. The significance of the braneworld theory is that the observable universe could be a 3+1 surface or brane, embedded in 3+1+d dimensional bulk spacetime. The particles predicted by the standard
	particle physics models and their corresponding fields may be trapped in the brane, while gravity can 
	access the higher dimensions of the bulk spacetime also. That is the information of the higher dimensional spacetime to our four dimensional world, is a subject matter of Braneworld models of gravity. As a result, it may be even possible to enquire about the influence of the higher dimensional bulk in causing the degrees of freedom on the horizon surface of the brane. For executing this extension we
	defined the surface degrees of freedom by using the relation of the horizon entropy in braneworld gravity theory. Further we compared the law of emergence thus obtained with the ad hoc form proposed by Sheykhi for braneworld models in reference\cite{sheykhi2013friedmann}.
	
	It is known that, every macroscopic system tends to evolve towards a state of thermodynamic equilibrium that aligns with its constraints \cite{H.B}. This in turn means that, over the time, the entropy of such systems reaches a maximum value. Pavon and Radicella \cite{Pav_n_2012} demonstrated that a Friedmann universe, with a Hubble expansion history, can behave like an ordinary macroscopic system, for which the horizon entropy eventually (which is the major of the entropy of the entire universe) approaching a maximum value. While in the context of the principle of emergence of cosmic space, 
	the universe’s evolution can be understood as being driven by the deviation from holographic equipartition, and the universe evolving towards a state that satisfies the holographic equipartition. Therefore it is natural to check the correspondance between the state of holographic equipartition and the state of maximum entropy. After obtaining the law of emergence from the unified first law of thermodynmics, we are
	analysisng, whether the law of emergence can lead to the maximisation of horizon entropy 
	in the braneworld scenarios.
	
	The paper is organized as follows. In section 2.1, we derive the law of emergence from the unified first law of thermodynamics for RS II braneworld. In section 2.2, we arrive at the law of emergence in Warped DGP braneworld model from the unified first law of thermodynamics. The same procedure is extended to the Gauss-Bonnet braneworld in section 2.3. We investigate horizon entropy maximization for the above models in section 3 and we present our conclusions in section 4.   
		
		\section{The law of emergence from the first law of thermodynamics in braneworld scenarios}
		In this section, we obtain the law of emergence from the unified first law of thermodynamics in the context of RS II braneworld, Warped DGP braneworld and Gauss-Bonnet braneworld.
		\subsection{The law of emergence from the first law of thermodynamics in RS II braneworld}
		
		
		An ad hoc form of the law of expansion in context of RS II braneworld cosmology was postulated in reference 
		\cite{sheykhi2013friedmann}, and have the form,
		\begin{equation}\label{el}
			\beta  \frac{dV}{dt} = G_{n+1}H\tilde{r}_A(N_{surf} - N_{bulk}),
		\end{equation}
		where, $\beta = {(n-2)}/({2(n-3)})$, $V$  is the volume of the horizon of the (n+1) dimensional universe, $t$ is the cosmic time and $G_{n+1}$ is the effective gravitational constant for the $n+1$ dimensional spacetime. The authors have suitably defined both $N_{surf}$ and $N_{bulk}$ using the definitions of entropy and Komar energy respectively. The arbitrariness in the proposal of this law, or in the form of the surface degrees of freedom, can be removed, if one could obtain these from a more fundamental principle. First law of thermodynamics is indeed such a fundamental principle. Hence, deriving this law 
		of emergence directly from the modified first law of thermodynamics will further motivate its fundamental physical nature.

		The unified first law of thermodynamics is given by\cite{PhysRevD.75.084003},
		 \begin{equation}\label{eqn:law1}
		 dE  =  TdS + WdV.
		 \end{equation} 
		  Here $E$ is the horizon energy, $T$ is the temperature of the horizon, $S$ is its entropy and the term $W=(\rho-p)/2,$ is the work density, which becomes,$-pdV$ for the de Sitter universe, for which $p=-\rho.$
		In RS II braneworld model, entropy of the apparent horizon for an $(n-1)$ brane embedded in $(n+1)$ dimensional bulk is given by (here $n$ is the spacial dimension),
		\begin{equation}\label{entropy}
			S = \frac{(n-1)l\Omega_{n-1}}{2G_{n+1}}\int_{0}^{\tilde{r}_A}\frac{\tilde{r}_A^{n-2}}{\sqrt{\tilde{r}_A^2 + l^2}},
		\end{equation}
		where, $l$ is the bulk AdS radius given by,
		\begin{equation}
			l^{2} = - \frac{n(n-1)}{16 \pi G_{n+1} \Lambda_{n+1}} 
		\end{equation}
		and $\tilde{r}_A$ is the apparent horizon radius of the FRW universe, given by,
		\begin{equation}
			\tilde{r}_A  = \frac{1}{\sqrt{H^2 + k/a^2}}
		\end{equation}
		where $H$ is the Hubble parameter, $a$ is the scale factor and $k$ is the curvature parameter of the universe. The term, $\Omega_{n-1}$ is the $(n-1)$ dimensional (spacial) volume given by the expression,
		\begin{equation}
			\Omega_{n-1}  = \frac{\pi^{(n-1)/2}}{\Gamma ((n + 1)/2)}.
		\end{equation}
		Using the idea of the  area law of entropy, the entropy is taken to be $S=A/4 G,$ where $A$ is the area of the horizon. Following this basic law, the entropy in the braneworld scenario, can be expressed in a similar form by defining an effective area as
		\cite{sheykhi2013friedmann},\\
		\begin{equation}
			\tilde{A} = 4G_{n+1}S = 2(n-1)l\Omega_{n-1}\int_{0}^{\tilde{r}_A}\frac{\tilde{r}_A^{n-2}}{\sqrt{\tilde{r}_A^2 + l^2}}.
		\end{equation}
		Then the corresponding effective rate of increase in the volume is \\
		\begin{equation}\label{rsdv}
			\frac{dV}{dt}  = \frac{\tilde{r}_A}{(n-2)}\frac{d\tilde{A}}{dt}
			=-2\Omega_{n-1}\frac{(n-1)}{(n-2)}\tilde{r}_A^{n+1}\frac{d}{dt}\Big(\sqrt{\tilde{r}_A^{-2} + \frac{1}{l^2}}\Big).
			\end{equation}
		
		Now to formulate the modified first law of thermodynamics, we need to have `$dE$' the change in energy and $dS$  the corresponding change in horizon entropy respectively. Change in the energy enclosed by the horizon can be obtained by considering the total gravitational energy inside volume, the Komar energy, which is given by,
		\begin{equation}
			E = \Omega_{n-1}\tilde{r}_A^{n-1}\rho.
		\end{equation}   
		Then the corresponding change in the volume energy is,  
		\begin{equation} \label{DE}
			dE = \Omega_{n-1}\tilde{r}_A^{n-1}d\rho + (n-1) \Omega_{n-1}\tilde{r}^{n-2}\rho d\tilde{r}_A.
		\end{equation}
		The change in the horizon entropy $dS,$ can be obtained using equation (\ref{entropy}), as \,
		\begin{flalign}
			dS &= \frac{(n-1)l\Omega_{n-1}}{2G_{n+1}}\frac{\tilde{r}_A^{n-2}}{\sqrt{\tilde{r}_A^2 + l^2}}d\tilde{r}_A.
		\end{flalign}
		Generally, the temperature of the horizon in a curved universe is a measure of its surface gravity and is given by\cite{Akbar_2007},
		\begin{flalign}
			T = \frac{\kappa}{2\pi} = \frac{1}{2\pi}\left[-\frac{1}{\tilde{r}_A}\left(1 - \frac{\dot{\tilde{r}}_A}{2H\tilde{r}_A}\right)\right],
		\end{flalign}
		where $\kappa$ is the surface gravity and the over-dot represents the derivative with respect to cosmic time.
		The term involving work density can be obtained as
		\begin{flalign}\label{wdV}
			WdV = \frac{(\rho - p)}{2}(n-1)\Omega_{n-1}\tilde{r}_A^{n-2}d\tilde{r}_A.
		\end{flalign}
		Using all the above results, the modified first law of thermodynamics in the equation (\ref{eqn:law1}) can now be expressed as, 
		\begin{multline}
			\Omega_{n-1}\tilde{r}_A^{n-1}d\rho + (n-1) \Omega_{n-1}\tilde{r}^{n-2}\rho d\tilde{r}_A  = \frac{1}{2\pi}\big[-\frac{1}{\tilde{r}_A}\big(1 - \frac{\dot{\tilde{r}}_A}{2H\tilde{r}_A}\Big)\Big]\frac{(n-1)l\Omega_{n-1}}{2G_{n+1}}\frac{\tilde{r}_A^{n-2}}{\sqrt{\tilde{r}_A^2 + l^2}}d\tilde{r}_A \\+ \frac{(\rho - p)}{2}(n-1)\Omega_{n-1}\tilde{r}_A
			^{n-2}d\tilde{r}_A.
		\end{multline}
		Let us rewrite the above equation in a convenient form as,
		\begin{multline*}
			\Omega_{n-1}\tilde{r}_A^{n-1}\dot\rho dt + (n-1) \Omega_{n-1}\tilde{r}^{n-2}\rho d\tilde{r}_A - \frac{(\rho - p)}{2}(n-1)\Omega_{n-1}\tilde{r}_A^{n-2}d\tilde{r}_A \\ =\frac{1}{2\pi}\big[-\frac{1}{\tilde{r}_A}\big(1 - \frac{\dot{\tilde{r}}_A}{2H\tilde{r}_A}\Big)\Big]\frac{(n-1)l\Omega_{n-1}}{2G_{n+1}}\frac{\tilde{r}_A^{n-2}}{\sqrt{\tilde{r}_A^2 + l^2}}d\tilde{r}_A.
		\end{multline*}               
		This equation can be further simplified using the respective Friedmann equation, 
		\begin{equation}\label{friedmann}
			\sqrt{H^2 + \frac{k}{a^2} + \frac{1}{l^2}} = \frac{4\pi G_{n+1}}{(n-1)}\rho.
			\end{equation}
		We then arrived at, 
		\begin{multline*}
			\frac{(n-1)\Omega_{n-1}\tilde{r}_A^{n-1}}{4\pi G_{n+1}}\Big.\frac{d}{dt}\Big(\big.\sqrt{\frac{1}{\tilde{r}_A^2} + \frac{1}{l^2}}\Big)dt + (n-1)\Omega_{n-1}\frac{(\rho + p)}{2}\tilde{r}_A^{n-2}d\tilde{r}_A\\=\frac{1}{2\pi}\big[-\frac{1}{\tilde{r}_A}\big(1 - \frac{\dot{\tilde{r}}_A}{2H\tilde{r}_A}\Big)\Big]\frac{(n-1)l\Omega_{n-1}}{2G_{n+1}}\frac{\tilde{r}_A^{n-2}}{\sqrt{\tilde{r}_A^2 + l^2}}d\tilde{r}_A. 
		\end{multline*}   
		This equation can be rewritten finally in to the form of the law of expansion. For that, we first  
		divide the above equation by $ dt $ and we simplify the R.H.S as,
		\begin{multline*}
			\frac{(n-1)\Omega_{n-1}\tilde{r}_A^{n-1}}{4\pi G_{n+1}}\Big.\frac{d}{dt}\Big(\big.\sqrt{\frac{1}{\tilde{r}_A^2} + \frac{1}{l^2}}\Big) + (n-1)\Omega_{n-1}\frac{(\rho + p)}{2}\tilde{r}_A^{n-2}   
			\,\dot{\tilde{r}}_A\ \\=\frac{1}{2\pi}\big[\frac{1}{\tilde{r}_A}\big(1 - \frac{\dot{\tilde{r}}_A}{2H\tilde{r}_A}\Big)\Big]\frac{(n-1)\Omega_{n-1}\tilde{r}_A^{n}}{2G_{n+1}}\Big.\frac{d}{dt}\Big(\big.\sqrt{\frac{1}{\tilde{r}_A^2} + \frac{1}{l^2}}\Big).
		\end{multline*}
Further simplification leads to
		\begin{equation}\label{eqn:mediumeq}
			(n-1)\Omega_{n-1}\frac{(\rho + p)}{2}\tilde{r}_A^{n-2}=\frac{-(n-1)\Omega_{n-1}\tilde{r}_A^{n-2}}{8\pi HG_{n+1}}\Big.\frac{d}{dt}\Big(\big.\sqrt{\frac{1}{\tilde{r}_A^2} + \frac{1}{l^2}}\Big).
		\end{equation}
		This equation can be suitably re-framed by identifying the rate of change of horizon volume,  $\frac{dV}{dt}=\frac{d}{dt}\left(\big.\sqrt{\frac{1}{\tilde{r}_A^2} + \frac{1}{l^2}}\right), as $ 
%
		\begin{equation}\label{e17}
			\beta \frac{dV}{dt}  =  H \tilde{r}_AG_{n+1}\Big[ N_{surf} - N_{bulk} \Big],
		\end{equation} \label{expansionlaw}
		where we have identified the surface degrees of freedom as,
		\begin{equation}
			N_{surf} = \frac{2\beta (n-1)\Omega_{n-1} }{(n-3)G_{n+1} }  \tilde{r}_A^{-2} \sqrt{ \tilde{r}_A^{-2}+\frac{1}{l^2}},
		\end{equation}
		and the degrees of freedom within the volume enclosed by the horizon as,
		\begin{equation}
			N_{bulk} = 4\pi \Omega_{n-1}\tilde{r}_A^n  \frac{(n-3)\rho + (n-1)p}{(n-3)}.
		\end{equation}
		Equation (\ref{e17}) is exactly similar to the proposed form of the expansion law given in equation (\ref{el}), but now we have arrived at it from a fundamental thermodynamic principle, the unified first law.		
		\subsection{The law of emergence from the first law of thermodynamics in Warped DGP braneworld}
	    In this section, we obtain the law of expansion from the unified first law of thermodynamics, considering another important braneworld gravity model, warped Dvali, Gabadadze and Porrati(DGP)braneworld. It is one of the extension of RS braneworld model which explains the late-time acceleration of the universe without any dark energy. In this model, gravity has two parts, one due to intrinsic curvature of the brane and other induced from $(n+1)$ dimensional bulk. Friedmann equation of this model is given in \cite{sheykhi2013friedmann}.
	    Now, consider $(n-1)$ dimensional warped DGP brane embedded in an $(n+1)$ dimensional AdS bulk. Entropy of apparent horizon in this model is given by,
	    \begin{equation}
	    	S = (n-1)\Omega_{n-1}\int_{0}^{\tilde{r}_A}\Big( \frac{(n-2)\tilde{r}_A^{n-3}}{4G_n} + \frac{l}{2G_{n+1}}  \frac{\tilde{r}_A^{n-2}}{\sqrt{\tilde{r}_A^2 + l^2}}\Big)d\tilde{r}_A.
	    \end{equation}
	    The effective surface area can then be defined as in \cite{sheykhi2013friedmann},
	    \begin{equation}
	    	\tilde{A} =4G_{n+1}S  =4G_{n+1}(n-1)\Omega_{n-1}\int_{0}^{\tilde{r}_A}\Big( \frac{(n-2)\tilde{r}_A^{n-3}}{4G_n} + \frac{l}{2G_{n+1}}  \frac{\tilde{r}_A^{n-2}}{\sqrt{\tilde{r}_A^2 + l^2}}\Big)d\tilde{r}_A.
	    \end{equation}
	    Then the infinitesimal rate of increase in effective volume can be expressed as \cite{sheykhi2013friedmann},
	    \begin{flalign}
	    	\frac{d\tilde{V}}{dt} = -2\Omega_{n-1}\frac{(n-1)}{(n-2)}\tilde{r}_A^{n+1}\Big.\frac{d}{dt}\Big(\frac{(n-2)G_{n+1}}{4G_n}\tilde{r}_A^{-2} + \sqrt{\tilde{r}_A^{-2} + \frac{1}{l^{2}}}\Big).
	    \end{flalign}
	    Using these equations the unified first law of thermodynamics can be written as,
	    \begin{multline*}
	    	\frac{1}{2\pi}\big[-\frac{1}{\tilde{r}_A}\big(1 - \frac{\dot{\tilde{r}}_A}{2H\tilde{r}_A}\Big)\Big](n-1)\Omega_{n-1}\Big[ \frac{(n-2)\tilde{r}_A^{n-3}}{4G_n} + \frac{l}{2G_{n+1}}  \frac{\tilde{r}_A^{n-2}}{\sqrt{\tilde{r}_A^2 + l^2}}\Big]d\tilde{r}_A \\ = -\Omega_{n-1}\tilde{r}_A^{n-1}(n-1)H(\rho + p)dt + (n-1)\Omega_{n-1}\frac{\rho + p}{2}\tilde{r}_A^{n-2}d\tilde{r}_A.
	    \end{multline*} 
	    Note that we have used equations (\ref{DE}) and (\ref{wdV}) for defining $dE$ and $WdV$ respectively because these are not model dependent. Now, with the help of continuity equation and Friedmann equation the  above expression can be simplified to
	    \begin{multline*}
	    	\frac{1}{2\pi}\big[-\frac{1}{\tilde{r}_A}\big(1 - \frac{\dot{\tilde{r}}_A}{2H\tilde{r}_A}\Big)\Big](n-1)\Omega_{n-1}\Big[ \frac{(n-2)\tilde{r}_A^{n-3}}{4G_n} + \frac{l}{2G_{n+1}}  \frac{\tilde{r}_A^{n-2}}{\sqrt{\tilde{r}_A^2 + l^2}}\Big]d\tilde{r}_A \\ = \frac{\Omega_{n-1}\tilde{r}_A^{n-1}(n-1)}{4\pi G_{n+1}}\Big.\frac{d}{dt}\Big(\frac{G_{n+1}}{4G_n}(n-2)\tilde{r}_A^{-2} + \sqrt{\tilde{r}_A^{-2} + \frac{1}{l^2}}\Big)dt + (n-1)\Omega_{n-1}\frac{\rho + p}{2}\tilde{r}_A^{n-2}d\tilde{r}_A.
	    \end{multline*} 
	    Further simplification leads to
	    \begin{equation}
	    	\frac{-(n-1)\Omega_{n-1}\tilde{r}_A^{n-2}}{8\pi G_{n+1}H}\Big.\frac{d}{dt}\Big( \frac{G_{n+1}}{4G_n}(n-2)\tilde{r}_A^{-2} + \sqrt{\tilde{r}_A^{-2} + \frac{1}{l^2}}\Big)  = (n-1)\Omega_{n-1}\frac{\rho + p}{2}\tilde{r}_A^{n-2}. 
	    \end{equation} 
	    Let us rewrite the above equation in a convenient form as,
	    \begin{multline*}
	    	\frac{-2(n-1)\Omega_{n-1}\tilde{r}_A^{n-2}}{2(n-3)}\Big.\frac{d}{dt}\Big( \frac{G_{n+1}}{4G_n}(n-2)\tilde{r}_A^{-2} + \sqrt{\tilde{r}_A^{-2} + \frac{1}{l^2}}\Big) \\ = 8\pi H\Omega_{n-1}G_{n+1}\Big( \frac{(n-3)\rho + (n-1)p}{2(n-3)} + \frac{\rho}{(n-3)}\Big)\tilde{r}_A^{n-2}.
	    \end{multline*} 
	    Simplifying the above equation, we get
	    \begin{multline*}
	    	\frac{-2(n-1)\Omega_{n-1}\tilde{r}_A^{n-2}}{2(n-3)}\Big.\frac{d}{dt}\Big( \frac{G_{n+1}}{4G_n}(n-2)\tilde{r}_A^{-2} + \sqrt{\tilde{r}_A^{-2} + \frac{1}{l^2}}\Big) \\ = HG_{n+1}\Big( 4\pi \Omega_{n-1}\tilde{r}_A^n\frac{(n-3)\rho + (n-1)p}{2(n-3)} +  8\pi \Omega_{n-1}\tilde{r}_A^n\frac{\rho}{(n-3)}\Big)\tilde{r}_A^{-2}.
	    \end{multline*} 
	    In the above equation identifying the relations for $N_{bulk}$ , $\beta$ and $\frac{dV}{dt}$, we arrive at
	    \begin{equation*}
	    	\beta \frac{dV}{dt} = H\tilde{r}_AG_{n+1}\Big( 8\pi \Omega_{n-1}\tilde{r}_A^n\frac{\rho}{(n-3)} - N_{bulk}\Big).
	    \end{equation*} 
	    Using Friedman equation this can be simplified to
	    \begin{equation*}
	    	\beta \frac{dV}{dt} = H\tilde{r}_AG_{n+1}\Big( N_{surf} - N_{bulk}\Big),
	    \end{equation*}
	    where
	    \begin{equation}
	    	N_{surf} = \frac{2\Omega_{n-1}(n-1)}{G_{n+1}(n-3)}\tilde{r}_A^{n}\Big(\frac{(n-2)G_{n+1}}{4G_n}\tilde{r}_A^{-2} + \sqrt{\tilde{r}_A^{-2} + \frac{1}{l^{2}}}\Big),
	    \end{equation}
	    and
	    \begin{equation}
	    	N_{bulk} = 4\pi \Omega_{n-1}\tilde{r_A^n}  \frac{(n-3)\rho + (n-1)p}{(n-3)}.
	    \end{equation}
	    It is important to note that this derived law of emergence is identical to the one proposed earlier.
\subsection{The law of emergence from the first law of thermodynamics in Gauss-Bonnet braneworld}
Here, we obtain 
the emergence principle 
in Guass-Bonnet braneworld model from the unified first law of thermodynamics. This model
is an extension of RS braneworld which can be obtained by adding a Gauss-Bonnet term to the gravity action. The resulting cosmological model is 5 dimensional 
with a 4 dimensional brane. Correspondingly, $n$ is taken to be equal to 4. 
Entropy of the apparent horizon of the universe in this model is given by \cite{sheykhi2013friedmann},
\begin{equation}
	S = \frac{3\Omega_3}{2G_4}\int_{0}^{\tilde{r}_A} \tilde{r}_A d\tilde{r}_A + \frac{3\Omega_3}{2G_5}\int_{0}^{\tilde{r}_A} \frac{\tilde{r}_A^2}{\sqrt{1-\Phi_0\tilde{r}_A^2}}d\tilde{r}_A + \frac{6\alpha\Omega_3}{G_5}\int_{0}^{\tilde{r}_A} \frac{2 - \Phi_0\tilde{r}_A^2}{\sqrt{1-\Phi_0\tilde{r}_A^2}}d\tilde{r}_A.
\end{equation} 
Accordingly, we can define the effective area of the apparent horizon as,
\begin{equation}
	\tilde{A} = 4G_5S  =  \frac{6G_5\Omega_3}{G_4}\int_{0}^{\tilde{r}_A} \tilde{r}_A d\tilde{r}_A  +  \frac{6G_5\Omega_3}{G_5}\int_{0}^{\tilde{r}_A} \frac{\tilde{r}_A^2}{\sqrt{1-\Phi_0\tilde{r}_A^2}}d\tilde{r}_A  + 24\alpha\Omega_3\int_{0}^{\tilde{r}_A} \frac{2 - \Phi_0\tilde{r}_A^2}{\sqrt{1-\Phi_0\tilde{r}_A^2}}d\tilde{r}_A.
\end{equation}
The rate of increase in the corresponding effective volume is then obtained as,
\begin{equation}
\frac{d\tilde{V}}{dt} = \frac{\tilde{r}_A}{2}\frac{d\tilde{A}}{dt}  =  \frac{3G_5\Omega_3}{G_4} \tilde{r}_A^2 \,\dot{\tilde{r}}_A  + 3\Omega_3 \frac{\tilde{r}_A^2 \,\dot{\tilde{r}}_A }{\sqrt{\tilde{r}_A^{-2} -\Phi_0}} +  12\alpha\Omega_3 \frac{2 - \Phi_0\tilde{r}_A^2}{\sqrt{\tilde{r}_A^{-2} -\Phi_0}}\,\dot{\tilde{r}}_A.
\end{equation}
Substituting these equations, we can formulate the modified first law of thermodynamics as,
\begin{multline}
\frac{1}{2\pi}\Big[\frac{-1}{\tilde{r}_A}\Big( 1 - \frac{\dot{\tilde{r}}_A}{2H\tilde{r}_A}\Big)\Big]\Big[ \frac{3\Omega_3}{2G_4 }\tilde{r}_A  + \frac{3\Omega_3}{2G_5} \frac{\tilde{r}_A^2}{\sqrt{1-\Phi_0\tilde{r}_A^2}} + \frac{6\alpha\Omega_3}{G_5} \frac{2 - \Phi_0\tilde{r}_A^2}{\sqrt{1-\Phi_0\tilde{r}_A^2}}\Big]d\tilde{r}_A\\=  -3\Omega_{3}\tilde{r}_A^{3}H(\rho + p)dt + 3\Omega_{3}\frac{\rho + p}{2}\tilde{r}_A^{2}d\tilde{r}_A,
\end{multline}
and with help of Friedmann equation we can rewrite the above equation in to more simple form,
\begin{equation*}
\frac{\dot{\tilde{r}}_A}{4 \pi H\tilde{r}_A^2}\Big[ \frac{3\Omega_3}{2G_4 }\tilde{r}_A  + \frac{3\Omega_3}{2G_5} \frac{\tilde{r}_A^2}{\sqrt{1-\Phi_0\tilde{r}_A^2}} + \frac{6\alpha\Omega_3}{G_5} \frac{2 - \Phi_0\tilde{r}_A^2}{\sqrt{1-\Phi_0\tilde{r}_A^2}}\Big]=  3\Omega_{3}\frac{\rho + p}{2}\tilde{r}_A^{2}.
\end{equation*}
Multiplying the above equation with $ 8\pi H \tilde{r}_A^3G_5$ and with necessary simplification, we can rewrite it as the conventional form of the law of emergence,
\begin{equation*}
\frac{dV}{dt} = G_5 H  \tilde{r}_A\Big(N_{surf} - N_{bulk}\Big).
\end{equation*}
Here, we identified the degrees of freedom as, 
\begin{equation}
N_{surf}  =  \frac{3\Omega_3}{G_4} \tilde{r}_A^2  + \frac{6\Omega_3}{G_5} \tilde{r}_A^4 \sqrt{\tilde{r}_A^{-2} -\Phi_0} +  \frac{16\alpha\Omega_3}{G_5}\tilde{r}_A^4 \Big(\tilde{r}_A^{-2} -\frac{\Phi_0}{2} \Big)\sqrt{\tilde{r}_A^{-2} -\Phi_0},
\end{equation}
and
\begin{equation}
N_{bulk} = 4\pi \Omega_{n-1}\tilde{r_A^n}  \frac{(n-3)\rho + (n-1)p}{(n-3)}.
\end{equation}
It has to be mentioned that the derived form of the law of emergence in Guass-Bonnet braneworld with the defined degrees of freedom is identical to the one, which is proposed in \cite{sheykhi2013friedmann}.
\section{The law of emergence and the horizon entropy maximization}
Any thermodynamic system proceed towards a maximum entropy state if
\begin{equation}
	\dot{S} \ge  0
\end{equation}
and
\begin{equation}
	\ddot{S} < 0, \, \, \textrm{at least at the end stage of the evolution. }
\end{equation}
In this section, we are going to see whether these conditions for entropy maximization hold in braneworld gravity models.
\subsection{The law of emergence and the horizon entropy maximization in RSII braneworld}
Here, we analyze whether the horizon entropy of the universe is getting maximized in the context of RSII braneworld exploring it's connection with the law of emergence. The derivative of entropy with respect to time can be expressed as,
\begin{equation} \label{dsdt}
	\frac{dS}{dt} = \frac{(n-1)l\Omega_{n-1}}{2G_{n+1}}\frac{\tilde{r}_A^{n-2}}{\sqrt{\tilde{r}_A^2 + l^2}} \dot{\tilde{r}}_{A},
\end{equation}
and from equation (\ref{rsdv}), we get
\begin{equation} \label{dvdt}
	\frac{dV}{dt} = -2\Omega_{n-1}l\frac{(n-1)}{(n-2)}\frac{\tilde{r}_A^{n-1}}{\sqrt{\tilde{r}_A^{2} + l^2}}\dot{\tilde{r}}_{A}.
\end{equation}
Comparing equations (\ref{dsdt}) and (\ref{dvdt}) we obtain,
\begin{equation}
	\frac{dV}{dt} = \frac{4G_{n+1}\tilde{r}_{A}}{(n-2)}\frac{dS}{dt}.
\end{equation}
Using the expansion law in the braneworld models, the above equation can be rewritten as,
\begin{equation}\label{lw}
	\beta\frac{4G_{n+1}\tilde{r}_{A}}{(n-2)}\frac{dS}{dt} = G_{n+1}H\tilde{r}_A(N_{surf} - N_{bulk}).
\end{equation} 
Re-arranging the above equation we get
\begin{equation}
	\frac{dS}{dt} = \frac{H(n-3)}{2}(N_{surf} - N_{bulk}).
\end{equation} 
Since  $N_{surf} \ge N_{bulk}$  as per the law of emergence, $\frac{dS}{dt} \ge 0$ throughout the evolution of the universe. Now calculating the second derivative of entropy, we get
\begin{equation}
	\ddot{S} = \frac{(n-3)\dot{H}}{2}(N_{surf} - N_{bulk}) + \frac{(n-3)H }{2}\frac{d}{dt}(N_{surf} - N_{bulk}).
\end{equation}
Here, first term will vanish in the final state, when $N_{surf} \rightarrow N_{bulk}$ in the long run. Also, as the universe is trying to decrease the holographic discrepancy, $\frac{d}{dt}(N_{surf}-N_{bulk})$ is negative, ensuring the non positivity of $\ddot{S}.$

\subsection{The law of emergence and the horizon entropy maximization in Warped DGP braneworld}
Here, we start with the rate of change of horizon entropy in Warped DGP braneworld, 
\begin{align}
	\frac{dS}{dt} &= (n-1)\Omega_{n-1}\Big[ \frac{(n-2)\tilde{r}_A^{n-3}}{4G_n} + \frac{l}{2G_{n+1}}  \frac{\tilde{r}_A^{n-2}}{\sqrt{\tilde{r}_A^2 + l^2}}\Big]\dot{\tilde{r}}_A\\
	             &= -\frac{(n-1)\Omega_{n-1}\tilde{r}_A^n}{ 2G_{n+1}}.\frac{d}{dt}\Big[ \frac{G_{n+1}}{4G_n}(n-2)\tilde{r}_A^{-2} + \sqrt{\tilde{r}_A^{-2} + \frac{1}{l^2}}\Big]\label{dsdt2}.
\end{align} 
Now, the rate of emergence can be expressed as,
\begin{equation} \label{dvdt2}
	\frac{dV}{dt} =  -2\Omega_{n-1}\frac{(n-1)}{(n-2)}\tilde{r}_A^{n+1}\Big.\frac{d}{dt}\Big(\frac{(n-2)G_{n+1}}{4G_n}\tilde{r}_A^{-2} + \sqrt{\tilde{r}_A^{-2} + \frac{1}{l^{2}}}\Big).
\end{equation}
Then, from equations (\ref{dsdt2}) and (\ref{dvdt2}) we obtain,
\begin{equation} 
	\frac{dV}{dt} =  4\frac{\tilde{r}_AG_{n+1}}{(n-2)}\dot{S}.
\end{equation}
This equation can be rewritten using expansion law in braneworld model as
\begin{equation}
	\beta\frac{4G_{n+1}\tilde{r}_{A}}{(n-2)}\dot{S} = G_{n+1}H\tilde{r}_A(N_{surf} - N_{bulk}).
\end{equation}
Re-arranging the above equation, we arrive at, 
\begin{equation}
	\frac{dS}{dt} = \frac{H(n-3)}{2}(N_{surf} - N_{bulk}).
\end{equation} 
Since  $N_{surf} \ge N_{bulk}$ as per the holographic equipartition, the horizon entropy in Warped DGP braneworld model never decreases. Differentiating the above equation, one gets 
\begin{equation}
	\ddot{S} = \frac{(n-3)\dot{H}}{2}(N_{surf} - N_{bulk}) + \frac{(n-3)H}{2}\frac{d}{dt}(N_{surf} - N_{bulk}).
\end{equation}
Here the first term will vanish in the long run, since the difference between the surface degrees of freedom and the bulk degrees of freedom is zero in the final state. Then in the long run $\ddot{S}$ will be
\begin{equation}
	\ddot{S} =  \frac{(n-3)H}{2}\frac{d}{dt}(N_{surf} - N_{bulk}).
\end{equation}
From the above expression it is clear that $\ddot{S} < 0$ in the final state of the universe, since the universe is trying to reduce the holographic discrepancy. Thus, in the context of Warped DGP braneworld the horizon entropy of the universe will never grow unbounded

\subsection{The law of emergence and the horizon entropy maximization in Gauss-Bonnet braneworld}
The rate of change of horizon entropy in the Gauss-Bonnet braneworld can be written as,

\begin{equation}\label{dsdt3}
	\frac{dS}{dt} = \frac{3\Omega_3}{2G_4} \tilde{r}_A \dot{\tilde{r}}_A + \frac{3\Omega_3}{2G_5} \frac{\tilde{r}_A^2}{\sqrt{1-\Phi_0\tilde{r}_A^2}}\dot{\tilde{r}}_A + \frac{6\alpha\Omega_3}{G_5} \frac{2 - \Phi_0\tilde{r}_A^2}{\sqrt{1-\Phi_0\tilde{r}_A^2}}\dot{\tilde{r}}_A.
\end{equation} 
Now, the rate change of effective volume is
\begin{align}
	\frac{d{V}}{dt} &= \frac{3G_5\Omega_3}{G_4} \tilde{r}_A^2 \,\dot{\tilde{r}}_A  + 3\Omega_3 \frac{\tilde{r}_A^2 \,\dot{\tilde{r}}_A }{\sqrt{\tilde{r}_A^{-2} -\Phi_0}} +  12\alpha\Omega_3 \frac{2 - \Phi_0\tilde{r}_A^2}{\sqrt{\tilde{r}_A^{-2} -\Phi_0}}\,\dot{\tilde{r}}_A.\\
	&= \tilde{r}_A\Big[\frac{3G_5\Omega_3}{G_4} \tilde{r}_A \,\dot{\tilde{r}}_A  + 3\Omega_3 \frac{\tilde{r}_A^2 \,\dot{\tilde{r}}_A }{\sqrt{1 -\Phi_0 \tilde{r}_A^2}} +  12\alpha\Omega_3 \frac{2 - \Phi_0\tilde{r}_A^2}{\sqrt{1 -\Phi_0 \tilde{r}_A^2}}\,\dot{\tilde{r}}_A.\Big]\label{dvdt3}
\end{align}
From equations (\ref{dsdt3}) and (\ref{dvdt3}) one can write,

\begin{equation}
	\frac{dV}{dt} = 2\tilde{r}_A G_{5} \dot{S}
\end{equation}
This equation can be rewritten using the expansion law in braneworld model as,
\begin{equation}
	2\beta G_{5}\tilde{r}_{A}\dot{S} = G_{5}H\tilde{r}_A(N_{suf} - N_{bulk}).
\end{equation}
Re-arranging the above equation, we get,
\begin{equation}
	\dot{S} = H(N_{suf} - N_{bulk}).
\end{equation} 
Since  $N_{surf} \ge N_{bulk}$ as per the holographic equipartition, $\frac{dS}{dt} \ge 0$ throughout the evolution of the universe. Differentiating the above equation, one gets
\begin{equation}
	\ddot{S} = \dot{H}(N_{suf} - N_{bulk}) + H\frac{d}{dt}(N_{suf} - N_{bulk}).
\end{equation}
Here the first term will vanish in the long run, since $N_{surf}$ approaches $N_{bulk}$ in the final state. Then $\ddot{S}$ will be
\begin{equation}
	\ddot{S} =  H\frac{d}{dt}(N_{suf} - N_{bulk}).
\end{equation}
Now, since the universe is trying to decrease the holographic discrepancy $N_{surf} - N_{bulk}$, $\ddot{S}$ will be negative in the final state. Hence it is shown that the horizon entropy of the universe is getting maximized in the context of Gauss Bonnet braneworld.

		\section{Conclusion}
		Emergent space paradigm is an 
		unique way to understand the expansion of the universe. In this paradigm, as originally proposed by Padmanabhan \cite{padmanabhan2009physical,Padmanabhan:2009ry}, expansion of the universe is driven by the difference between the degrees of freedom on the horizon and that within the bulk enclosed by it. 
		This principle, which was first proposed in the context of Einstein's gravity, 
		has been extended to more general 
		gravity theories like Gauss-Bonnet and Lovelock by properly defining 
		the surface degree of freedom in the respective gravity models
		\cite{cai2004note,TU2018411,PhysRevD.87.061501}. 
		In all of these
		works, the authors have just proposed the law of emergence in an ad hoc manner and then ratified it by deriving the Friedmann equations from it. Hence, it becomes highly demanding that, whether the law of emergence can be derived from some more fundamental principles. 
		
		In this 
		paper, we derived the 
		emergence law 
		from the fundamental thermodynamic relation, $dE = TdS + W dV, $ the unified first law, in the context of Braneworld gravity models. Braneworld scenario is a more general theory of gravity, which considers the 3+1 universe as a brane embedded in higher dimensional bulk, such that the higher dimensional bulk can influence the brane through the gravitational interaction. 
		Hence it is also considered to be a potential theory of gravity, which can offer a solution to the long standing cosmological constant problem. Hence, it becomes highly recommending to formulate the law of emergence in Braneworld models, starting from the most fundamental principle of thermodynamics. We have obtained the law of emergence from the modified first law of thermodynamics, 
		for three cases of 
		braneworld models:
		(i) RS II braneworld, (ii) Warped DGP model and (iii) Gauss-Bonnet braneworld. This shows that, like in Einstein's gravity the law of emergence is perfectly valid for most of the Braneworld scenarios. It is to be mentioned that the law of emergence derived from the thermodynamic law is matching with the respective ad hoc forms of the law in the respective models of Braneworld. Further more, our results ratify the fact that, the unified first law turns out to be the back bone of the law of emergence.

		We further analyse the connection of the law of emergence with the evolution of the horizon entropy. Our analysis shows that the
			law of expansion in three Braneworld models are equivalent to the principle of maximization of horizon entropy. During it's expansion the universe tends to decrease the departure from the holographic equipartition as per the law of emergence which can be considered as a tendency for maximizing the horizon entropy. This in turn implies that the law of emergence effectively implies the maximization of horizon entropy in the braneworld scenarios.
			

		\section*{Acknowledgment}
		
		P. B. Krishna acknowledges Cochin University of Science and Technology for the Postdoctoral Fellowship.


\begin{thebibliography}{}
		\bibitem{Hawking} S.W Hawking, Commun.Math.Phys.43,199(1975)
		\url{http://dx.doi.org/10.1007/BF02345020}
		
		
		\bibitem{medved2004conceptual}Medved A. \& Vagenas E. When conceptual worlds collide: The generalized uncertainty principle and the Bekenstein-Hawking entropy. {\em Physical Review D—Particles, Fields, Gravitation, And Cosmology}. \textbf{70}, 124021 (2004)
		\url{https://journals.aps.org/prd/abstract/10.1103/PhysRevD.70.124021}
		
		\bibitem{bekenstein1973black}Bekenstein J. Black holes and entropy. {\em Physical Review D}. \textbf{7}, 2333 (1973)
		\url{https://doi.org/10.1103/PhysRevD.7.2333}
		
		
		
		\bibitem{bardeen}J.M.Bardeen, B.Carter and S.W.Hawking, The four laws of black hole mechanics, Commun.Math.Phys.31,161(1973)
		\url{https://doi.org/10.1007/BF01645742}
		
		
		\bibitem{eling2006nonequilibrium}Eling C., Guedens R. \& Jacobson T. Nonequilibrium thermodynamics of spacetime. {\em Physical Review Letters}. \textbf{96}, 121301 (2006)
		\url{https://doi.org/10.1103/PhysRevLett.96.121301}	
		
		\bibitem{jacob} T. Jacobson, Thermodynamics of space-time: the Einstein equation of state, Phys. Rev. Lett.
		\textbf{75} 1260(1995)
		\url{https://doi.org/10.1103/PhysRevLett.75.1260}
		
		\bibitem{thermodynical_aspects} T. Padmanabhan, Thermodynamical aspects of gravity: new insights, Rept. Prog. Phys. \textbf{73} 046901 (2010) 
		\url{https://doi.org/10.1088/0034-4885/73/4/046901}
		
		\bibitem{pad_cosmos} T. Padmanabhan, Do we really understand the cosmos?, Comptes Rendus Physique \textbf{18} 275-291(2017)
		\url{https://doi.org/10.1016/j.crhy.2017.02.001}
		
		
		\bibitem{Sindoni} L. Sindoni, Emergent models for gravity: an overview of microscopic models, SIGMA \textbf{8} 027(2012)
		\url{https://doi.org/10.3842/SIGMA.2012.027}
		
		
		\bibitem{Emergent_gravity} T. Padmanabhan, Emergent gravity paradigm: recent progress, Mod. Phys. Lett. A \textbf{30} 1540007(2015)
		\url{https://doi.org/10.1142/S0217732315400076}
		
		
		\bibitem{Verli}E.P. Verlinde, On the origin of gravity and the laws of Newton, JHEP \textbf{04} 029(2011) 
		\url{https://doi.org/10.1142/S0217732315400076}
		
		
		\bibitem{pad_Equi}T. Padmanabhan, Equipartition of energy in the horizon degrees of freedom and the emergence
		of gravity, Mod. Phys. Lett. A \textbf{25} 1129(2010)
		\url{https://doi.org/10.1142/S021773231003313X}
		
		\bibitem{padmanabhan2016exploringnaturegravity}Padmanabhan, T. Exploring the Nature of Gravity. (2016), 
		\url{https://doi.org/10.48550/arXiv.1602.01474}
		
		
		\bibitem{padmanabhan2016atoms}T. Padmanabhan, The atoms of space, gravity and the cosmological constant. {\em International Journal Of Modern Physics D}. \textbf{25}, 1630020 (2016)
		\url{https://doi.org/10.1142/S0218271816300202}
		
		\bibitem{cai2017gravitational}Cai R., Cao, Z. Guo, Z. Wang, S. \& Yang T. The gravitational-wave physics. {\em National Science Review}. \textbf{4}, 687-706 (2017)
		\url{https://doi.org/10.1093/nsr/nwx029}
		
		\bibitem{PhysRevD.88.067303}Eune M. \& Kim W. Emergent Friedmann equation from the evolution of cosmic space revisited. {\em Phys. Rev. D}. \textbf{88}, 067303 (2013,9)
		\url{https://link.aps.org/doi/10.1103/PhysRevD.88.067303} 
		
		
		\bibitem{PhysRevD.96.063513}P. B. Krishna \& T. K. Mathew, Holographic equipartition and the maximization of entropy. {\em Phys. Rev. D}. \textbf{96}, 063513 (2017,9)
		\url{https://link.aps.org/doi/10.1103/PhysRevD.96.063513}
		
		\bibitem{PhysRevD.99.023535} P. B. Krishna \& T. K. Mathew, Entropy maximization in the emergent gravity paradigm. {\em Phys. Rev. D}. \textbf{99}, 023535 (2019,1) 
		\url{https://link.aps.org/doi/10.1103/PhysRevD.99.023535}
		
		
		\bibitem{krishna2022emergence} P. B. Krishna, Hassan Basari, V. \& T. K. Mathew, Emergence of cosmic space and its connection with thermodynamic principles. {\em General Relativity And Gravitation}. \textbf{54}, 58 (2022)
		\url{https://doi.org/10.1007/s10714-022-02941-4}
		
		\bibitem{krishna2020does} P. B. Krishna \& T. K Mathew, Does holographic equipartition demand a pure cosmological constant?. {\em Modern Physics Letters A}. \textbf{35}, 2050334 (2020)
		\url{https://doi.org/10.1142/S0217732320503344}
		
		\bibitem{krishna2024emergence} P. B. Krishna, \& T. K. Mathew, Emergence of cosmic space and the maximization of horizon entropy. {\em Physics Of The Dark Universe}. \textbf{44} pp. 101451 (2024)
		\url{https://doi.org/10.48550/arXiv.2002.02121}		
		
		\bibitem{vt2022emergence}Hassan Basari V. T, P. B. Krishna, Priyesh K. \& T. K. Mathew, Emergence of space and expansion of Universe. {\em Classical and Quantum Gravity}. \textbf{39}, 115012 (2022)
		\url{https://doi.org/10.1088/1361-6382/ac6a39}
		
		\bibitem{hassan2023unified}Hassan Basari V. T, P. B. Krishna, \& T. K. Mathew, Unified formalism for the law of emergence from the first law of thermodynamics. {\em Physical Review D}. \textbf{107}, 063511 (2023)
		\url{https://doi.org/10.1103/PhysRevD.107.063511}
		
		
		\bibitem{sheykhi2009generalized}Sheykhi A. \& Wang B. Generalized second law of thermodynamics in Gauss–Bonnet braneworld. {\em Physics Letters B}. \textbf{678}, 434-437 (2009)
		\url{https://doi.org/10.1016/j.physletb.2009.06.075}
		
		\bibitem{sheykhi2007thermodynamical}Sheykhi A., Wang B. \& Cai R. Thermodynamical properties of apparent horizon in warped DGP braneworld. {\em Nuclear Physics B}. \textbf{779}, 1-12 (2007)
		\url{https://doi.org/10.1016/j.nuclphysb.2007.04.028}
		
		\bibitem{komatsu2017cosmological}Komatsu N. Cosmological model from the holographic equipartition law with a modified Rényi entropy. {\em The European Physical Journal C}. \textbf{77}, 229 (2017)
		\url{https://doi.org/10.1140/epjc/s10052-017-4800-2}
		
		
		\bibitem{komatsu2023evolution}Komatsu N. Evolution of thermodynamic quantities on cosmological horizon in $\lambda(t)$ model. {\em Physical Review D}. \textbf{108}, 083515 (2023)
		\url{https://doi.org/10.1103/PhysRevD.108.083515}
		
		
		\bibitem{komatsu2024cosmological}Komatsu N. Cosmological model based on both a holographiclike connection and Padmanabhan’s holographic equipartition law. {\em Physical Review D}. \textbf{109}, 023505 (2024)
		\url{https://doi.org/10.1103/PhysRevD.109.023505}
		
		\bibitem{heydarzade2016emergent}Heydarzade Y., Hadi H., Darabi F. \& Sheykhi, A. Emergent universe in the braneworld scenario. {\em The European Physical Journal C}. \textbf{76} , 1 (2016)
		\url{https://doi.org/10.1140/epjc/s10052-016-4162-1}
		
		\bibitem{volovick} Grigory E Volovik, Thermodynamics and Decay of de Sitter Vacuum. {\em Symmetry}. \textbf{16}, 763 (2024)
		\url{https://doi.org/10.3390/sym16060763}
		
		\bibitem{Mirza} Dezaki F. L, Mirza B., Generalized entropies and the expansion law of the universe. {\em General Relativity and Gravitation}. \textbf{47}, 67 (2015)
		\url{https://doi.org/10.1007/s10714-015-1910-8}
		
		\bibitem{mahith2018expansion}Mahith M., P. B. Krishna, \& T. K. Mathew, Expansion law from first law of thermodynamics. {\em Journal Of Cosmology And Astroparticle Physics}. \textbf{2018}, 042 (2018)
		\url{https://doi.org/10.1088/1475-7516/2018/12/042}
		
		\bibitem{hareesh2019first}Hareesh T., P. B. Krishna, \& T. K. Mathew, First law of thermodynamics and emergence of cosmic space in a non-flat universe. {\em Journal Of Cosmology And Astroparticle Physics}. \textbf{2019}, 024 (2019)
		\url{https://doi.org/10.1088/1475-7516/2019/12/024}
		
		\bibitem{sheykhi2013friedmann}Sheykhi A., Dehghani M. \& Hosseini S. Friedmann equations in braneworld scenarios from emergence of cosmic space. {\em Physics Letters B}. \textbf{726}, 23-27 (2013)
		\url{https://doi.org/10.1016/j.physletb.2013.08.035}
		
		\bibitem{H.B}H.B. Callen, Thermodynamics (Wiley, New York, 1960)	
		
		\bibitem{Pav_n_2012}Pavón D. \& Radicella N. Does the entropy of the Universe tend to a maximum?. {\em General Relativity And Gravitation}. \textbf{45}, 63-68 (2012,9)
		\url{http://dx.doi.org/10.1007/s10714-012-1457-x}
		
		\bibitem{Moradpour_2016}Moradpour H. Implications, Consequences and Interpretations of Generalized Entropy in the Cosmological Setups. {\em International Journal Of Theoretical Physics}. \textbf{55}, 4176-4184 (2016)
		\url{http://dx.doi.org/10.1007/s10773-016-3043-6}
		
		\bibitem{Chen2022}Chen, G. Emergence of cosmic space and horizon entropy maximization from Tsallis and Cirto entropy. {\em The European Physical Journal C}. \textbf{82}, 532 (2022)
		\url{https://doi.org/10.1140/epjc/s10052-022-10474-y}
		
		
		\bibitem{PhysRevD.75.084003}Akbar M. \& Cai R. Thermodynamic behavior of the Friedmann equation at the apparent horizon of the FRW universe. {\em Phys. Rev. D}. \textbf{75}, 084003 (2007)
		\url{https://link.aps.org/doi/10.1103/PhysRevD.75.084003}
		
		\bibitem{Akbar_2007}Akbar M. \& Cai R. Thermodynamic behavior of field equations {\em Physics Letters B}. \textbf{648}, 243-248 (2007)
		\url{http://dx.doi.org/10.1016/j.physletb.2007.03.005}
		
		\bibitem{padmanabhan2009physical}Padmanabhan T. A physical interpretation of gravitational field equations. AIP Conf.Proc.1241:93-108,2010 (2009)
		\url{https://doi.org/10.1063/1.3462738}
		
		\bibitem{Padmanabhan:2009ry}Padmanabhan T. Entropy density of spacetime and thermodynamic interpretation of field equations of gravity in any diffeomorphism invariant theory.  (2009)
		\url{https://doi.org/10.48550/arXiv.0903.1254}
		
		\bibitem{cai2004note}Cai R. A note on thermodynamics of black holes in Lovelock gravity. {\em Physics Letters B}. \textbf{582}, 237-242 (2004)
		\url{https://doi.org/10.1016/j.physletb.2004.01.015}
		
		\bibitem{TU2018411}Tu F., Chen Y., Sun B. \& Yang Y. Accelerated expansion of the universe based on emergence of space and thermodynamics of the horizon. {\em Physics Letters B}. \textbf{784}  411 (2018)
		\url{https://doi.org/10.1016/j.physletb.2018.08.030}
		
		\bibitem{PhysRevD.87.061501}Sheykhi A. Friedmann equations from emergence of cosmic space. {\em Phys. Rev. D}. \textbf{87}, 061501 (2013)
		\url{https://link.aps.org/doi/10.1103/PhysRevD.87.061501}
		
		
		\bibitem{maartens2010brane}Maartens R. \& Koyama K. Brane-world gravity. {\em Living Reviews In Relativity}. \textbf{13} 5 (2010)
		\url{https://doi.org/10.12942/lrr-2010-5}	
		
	\end{thebibliography}
	\end{document}